\documentclass[iopartl,showpacs,preprintnumbers,amsmath]{revtex4}

\usepackage{siunitx}
\usepackage{amsmath}
\usepackage{graphicx}
\usepackage{color}

\begin{document}
\title{Stick-jump mode in surface droplet dissolution}

\author{Erik Dietrich$^{1,2}$}
\email{e.dietrich@utwente.nl}
\author{E. Stefan Kooij$^1$}
\author{Xuehua Zhang$^{3,2}$}
\author{Harold J. W. Zandvliet$^1$}
\email{h.j.w.zandvliet@utwente.nl}
\author{Detlef Lohse$^{2}$}
\email{d.lohse@utwente.nl}
\affiliation{$^1$Physics of Interfaces and Nanomaterials,  and $^2$Physics of Fluids, MESA+ Institute for Nanotechnology, University of Twente, P.O. Box 217, 7500 AE Enschede, The Netherlands. $^3$School of Civil, Environmental and Chemical Engineering, RMIT University, Melbourne, VIC 3001, Australia}

\begin{abstract}
The analogy between evaporating surface droplets in air to dissolving long-chain alcohol droplets in water is worked out. We show that next to the three known modi for surface droplet \textcolor{black}{evaporation or dissolution} (constant contact angle mode, constant contact radius mode, and stick-slide mode), a fourth mode exists for small droplets on supposedly smooth substrates, namely the stick-jump mode: intermittent contact line pinning causes the droplet to switch between sticking and jumping during the dissolution. We present experimental data and compare them to theory to predict the dissolution time in this stick-jump mode. We also explain why these jumps were easily observed for microscale droplets but not for larger droplets.
\end{abstract}

\maketitle

\section{Introduction}
The evaporation of sessile droplets \textcolor{black}{in air} is of key importance for inkjet printing, surface coating, cleaning, and the deposition of small particles, see e.g. \cite{Deegan2000,Maheshwari2008,Popov2005,Cazabat2010,Marin2011}. As a result, a great body of theoretical, numerical, and experimental work has been done in this research area, see e.g. \cite{Gelderblom2011,Erbil2012,Jansen2013}. A less studied, but completely analogous system is that of a \textcolor{black}{liquid} droplet surrounded by another liquid. Under certain conditions, namely when both processes are purely diffusion controlled, the two systems can be described by the same equations. In both cases, pinning drastically alters the evaporation or dissolution behavior of the droplet \cite{Zhang2015}, changing the dissolution rates and the motion of the droplet \cite{Hu2002, Nguyen2012}.

Picknett and Bexon \cite{Picknett1977} described two basic modi in which surface droplets can \textcolor{black}{evaporate}. In the first mode, the contact line of the droplet is pinned to the substrate, and the droplet wets the very same area throughout the \textcolor{black}{evaporation} process. The \textcolor{black}{base} diameter $L$ of the drop remains constant, hence the name of this mode: Constant Radius (CR). In the second process, the contact line is free to slide over the substrate so that the droplet can maintain the same contact angle $\theta$ during  \textcolor{black}{evaporation}. This mode is called the Constant Contact Angle mode (CA). 
Recently, Stauber and coworkers \cite{Stauber2014} described a third mode which they called 'stick-slide' mode: droplets evaporated initially in the CR mode till a certain contact angle $\theta^*$ was reached. From this point on, the \textcolor{black}{evaporation} mode switched to the CA mode (with contact angle $\theta^*$). The CA mode could last until the drop had disappeared, or could be followed by one or more subsequent stick-slide cycles. In the work of Stauber {\it{et al.}}, the duration of the 'stick' period was comparable to the 'slide' period.

\begin{figure}
\begin{center}
\includegraphics[angle=0,width=12cm]{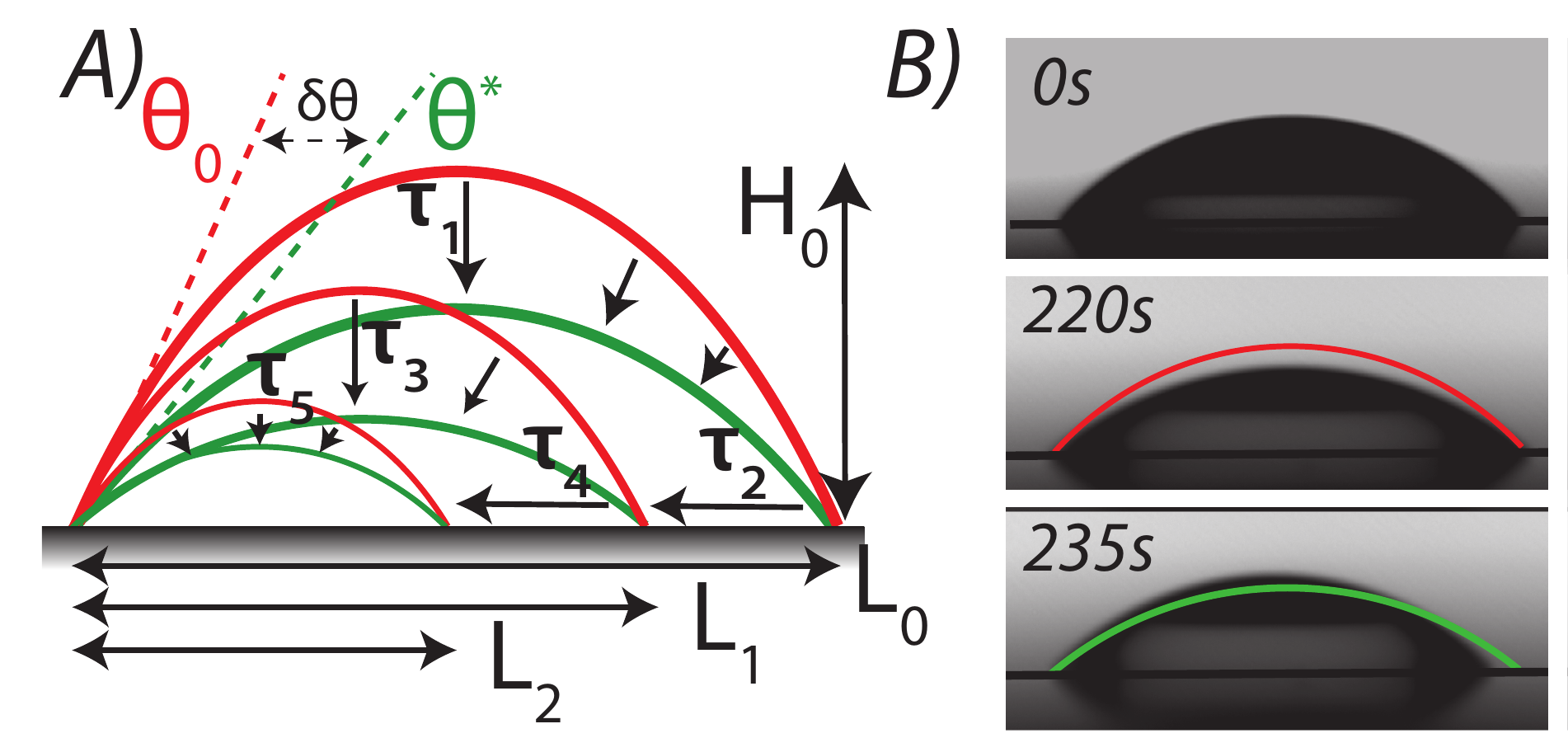}
\end{center}
\caption{\label{fig:sketch} (color online) Illustration (A) of the stick-jump dissolution mode. A water immersed 3-heptanol droplet on hydrophobized silicon with initial height $H_0$, \textcolor{black}{base} diameter $L_0$ and contact angle $\theta_0$ dissolves in the CR mode during time interval $\tau_1$. When $\theta=\theta^*$, the droplet jumps to retrieve $\theta_0$  and a smaller \textcolor{black}{base} $L_1$. This process is repeated with dissolution in the CR mode during time interval $\tau_3$, a jump during time interval $\tau_4\ll\tau_3$, and dissolution during $\tau_5$ and so on, until complete dissolution. Figure (B) shows snapshots of the initial droplet at $t=0$ s, where the black line indicates the silicon surface. After a total time of $t=220$ s  the droplet contact angle has decreased till $\theta^*$. The red line indicates the initial shape of the droplet. At $t=235$ s, the droplet has jumped to a new geometry and contact angle $\theta_0$. The outline of the droplet before the jump is added in green for comparison.}
\end{figure}

\textcolor{black}{In this paper we will study the 'stick-jump' mode in the context of droplet dissolution. In this mode, sketched in Figure \ref{fig:sketch}, a droplet has an initial base diameter $L_0$ and initial contact angle $\theta_0$.} During time interval $\tau_1$, the droplet is pinned and dissolves in the CR mode until the contact angle reaches $\theta^*$. In the next time interval $\tau_2$, the droplet `jumps' to a new geometry with contact angle $\theta_0$, and \textcolor{black}{base} diameter $L_1$. This process continues with dissolution in the CR mode during another time interval $\tau_3$, and a jump during interval $\tau_4\ll\tau_3$.
This mode has been shown to occur in evaporating droplets containing substantial amounts of particles \cite{Orejon2011, Maheshwari2008}, on specially patterned substrates \cite{Debuisson2011a,Debuisson2011b} and incidentally in other systems \cite{Mohammad2005,Bourges1995} where it also has been referred to as the `stick-slip' mode. \textcolor{black}{To avoid confusion between slip and slide, we will use the term "stick-jump" instead of "stick-slip" to describe the mode in which $\tau_4\ll\tau_3$. We would like to emphasize that "jump" does not mean that the droplet detaches from the substrate, it merely reflects the droplets abrupt change in geometry, i.e., an increase in its height $H$.}  

In this paper we will demonstrate that the stick-jump mode can occur for \textcolor{black}{liquid} droplets on un-patterned substrates.
The paper will start with a description of the experiment and the experimental results. Subsequently, a theoretical model is developed, to which the experimental results are compared. This comparison will highlight the important parameters in the stick-jump dissolution mode.

%%%FIGURE%%%%%%%%%%%%%%%%%%%%%%%%%%%%%%%%%%%%%%%%%%
%%%FIGURE%%%%%%%%%%%%%%%%%%%%%%%%%%%%%%%%%%%%%%%%%%

\section{Experimental procedure}
The droplet liquid in this study was 3-heptanol (Aldrich, 99\% purity and used as received). This long-chain alcohol has a solubility in water between  $4.0$ kg m$^{-3}$ at $25^{\circ}\mathrm{C}$ and $4.7$ kg m$^{-3}$ at $20^{\circ}\mathrm{C}$ \cite{barton1984,yalkowsky2010handbook}, and a density of $\rho=819\,$kg$\,$m$^{-3}$. Since for 3-heptanol no experimental values could be found for the diffusion constant and interfacial tension with water, the values for 1-heptanol are adopted: $D \approx 0.8\times10^{-9}$ m$^2$s$^{-1}$\textcolor{black}{)\cite{Hao1996}} and $\gamma = 7.7$ mN m$^{-1}$ \cite{Demond1993}.
Polished silicon wafers (P/Boron/(100), Okmetic) and glass microscope slides (Menzel Gl\"aser) were cut into small pieces of about 1 cm$^2$ and cleaned in a hot Piranha bath followed by a SC-1 bath (H$_2$O:H$_2$O$_2$:NH$_4$OH mixture)  and a SC-2 bath (H$_2$O:H$_2$O$_2$:HCl mixture) \cite{Karptischkathesis}. In between each step, the samples were \textcolor{black}{rinsed} with MilliQ water (obtained from a Reference A+ system, Merck Millipore, at $18.2$ M$\Omega$ cm) and insonicated in hot MilliQ water for 15 minutes.
After cleaning, the samples were coated via chemical vapor deposition with PFDTS (1H,1H,2H,2H-Perfluorodecyltrichlorosilane 97\%, ABCR Gmbh, Karlsruhe Germany). The samples were then annealed for 1 hour in an oven at $100^{\circ}\mathrm{C}$ and sonicated in chloroform for $\approx 10$ minutes. The surfaces were characterized with atomic force microscopy (Agilent 5100) in tapping mode using NSC36 cantilevers (MikroMasch).
The samples were cleaned in chloroform (Emsure $\ge99\%$ purity, Merck) in an ultrasound bath for 10 minutes, prior to each measurement. 
The sample was placed in the center of a square acrylic container and $350$ ml of MilliQ water was carefully added. This water was stored in a clean glass flask for a few hours prior to the experiment to reach room temperature. 
A small droplet, typically $10-30$ nL of 3-heptanol, was subsequently placed on the substrate using a glass syringe mounted in a motorized syringe pump. 
Imaging of the droplet was done through two CCD-cameras, mounted with long distance microscopes. The first camera imaged the front view of the droplet, while the second camera recorded the side view (for silicon substrates) or, via a mirror, the bottom view of the droplet on the glass substrate. Using two axis imaging allowed us to track the position of the drop, and to confirm that the droplet maintained its spherical cap shape during dissolution. 
The acquired pictures were post-processed in a Matlab program, which traced the droplet shape with sub-pixel accuracy \cite{vandermeulen2014}. This shape was fitted with a spherical cap and the intersection of the circle with the \textcolor{black}{base} plane was measured to extract the contact angle and \textcolor{black}{base} diameter. Droplet height and radius of curvature could be extracted from these values using goniometric relations. Relevant parameters from both observation axes were compared and interpolated.

\section{Experimental Results}

\begin{figure}
\begin{center}
\includegraphics[angle=0,width=17cm]{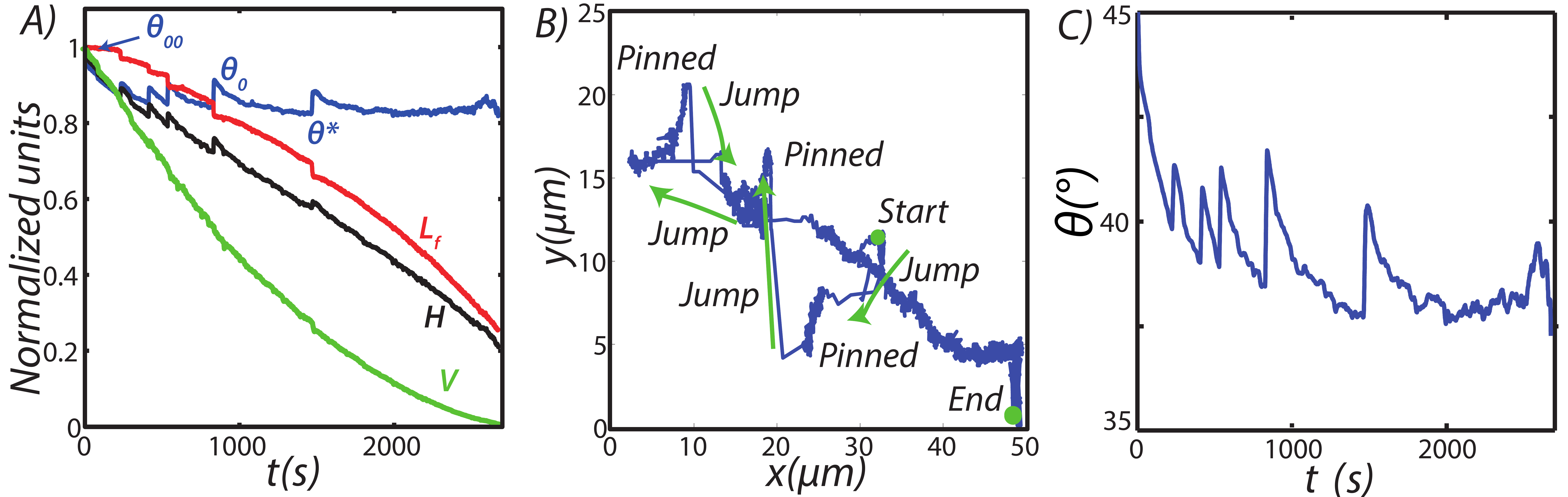}
\end{center}
\caption{\label{fig:Fig1} (color online) Dissolution of a water immersed 3-heptanol droplet on hydrophobized silicon: (A) Normalized volume (green), \textcolor{black}{base} diameter (red), droplet height (black) and contact angle (blue) in time. To allow for an easy comparison, all parameters have been normalized to their initial value: $26$ nl, $680\,\mu$m, $138\,\mu$m and $44.2^{\circ}$, respectively. The graph is truncated after 2800 seconds, when only a small amount of residue was left on the substrate. (B) Location of the droplet center in the horizontal plane, during the dissolution process. The position of the droplet center shows little change during the pinned phases, and abruptly moves to a new position when the droplet depins. (C) Contact angle $\theta(t)$. }
\end{figure}

In Figure \ref{fig:Fig1}, the evolution of the parameters characterizing the droplet during dissolution is shown. At $t=0$ s, directly after deposition, the droplet had a large contact angle $\theta_{00}$ which decreased as the droplet dissolved. The decrease in contact angle, at fixed \textcolor{black}{base} diameter, resulted in a decreasing height. When the contact angle has decreased to $\theta^*<\theta_{00}$, the contact line depinned and the droplet changed its geometry: the \textcolor{black}{base} diameter 'jumped' to a new and smaller \textcolor{black}{base} diameter, which yielded an increased contact angle and height. Note that the volume was conserved during the jump and that the original contact angle $\theta_{00}$ was not fully recovered: the droplet jumped back to a new contact angle $\theta_0$, where $\theta_{00}>\theta_0>\theta^*$. Usually, one part of the contact line stayed pinned during the jump, causing the rest of the droplet to move towards this point. This is illustrated in Figure \ref{fig:Fig1}b, where the position of the droplet center is shown as it moved over the substrate.

%%%FIGURE%%%%%%%%%%%%%%%%%%%%%%%%%%%%%%%%%%%%%%%%%%
%%%FIGURE%%%%%%%%%%%%%%%%%%%%%%%%%%%%%%%%%%%%%%%%%%

From Figure \ref{fig:Fig1}c, it appears that the largest contact angle $\theta_{00}$ was observed at $t=0$ s. This was caused by contact angle hysteresis during the deposition of the droplet \textcolor{black}{\cite{Gao2006}}: A small droplet was put at the substrate and then inflated to the desired volume. For simplicity in description and notation we will omit the difference between $\theta_{00}$ and $\theta_0$. Unless explicitly stated otherwise, we will assume that the droplet starts at $\theta_0$, dissolves until $\theta^*$ and jumps to $\theta_0$, with a corresponding change in contact angle of $\delta\theta=\theta_0-\theta^*$. 

The exact values for $\theta^*$ and $\theta_0$ differ from jump to jump as seen from Figure \ref{fig:Fig1}c and the \textcolor{black}{base} diameter shows a slow decay in between the jumps. This last observation is consistent with the findings of Orejon {\it{et al.}} \cite{Orejon2011} for weakly pinned evaporating droplets. This is understandable, since the values for $\theta$ and the pinning depend on local heterogeneities which spatially vary. After a jump, the contact line will be pinned by numerous small pinning sites, each with an individual pinning strength. As the contact angle decreases, the contact line first depins from the 'weaker' sites, giving rise to a slow creep in the \textcolor{black}{base} radius. When the contact angle is reduced to $\theta^*$, the drop is only pinned by the strongest pinning sites. A further decrease in $\theta$ will result in the next jump. 

To verify the influence of surface roughness, we repeated the experiments on hydrophobized glass slides. The glass slides were coated using exactly the same procedure, but they exhibit a greater intrinsic roughness (AFM images and more details on the roughness are provided in appendix B).  The rougher surface led to stronger pinning, which is reflected in significantly altered contact angles. For comparison, the main results on silicon and on glass are summarized in Table \ref{table1}. The higher roughness on glass obviously increased the pinning strength, leading to higher initial contact angles $\theta_{00}$ and lower values for $\theta^*$ and $\theta_0$.

\textcolor{black}{The observation of stick-jump behavior can therefore be related to a spatial variation in surface roughness and pinning. It is to be expected that a uniform roughness will result in stick-slide behavior: once the contact angle has reached $\theta^*$ and started to move, it only encounters pinning sites of equal strength, corresponding to $\theta^*$. Dissolution will thus proceed in the CA mode.  A local variation in pinning sites will result in temporary pinning of the contact line by the strongest sites, followed by the jump during which the contact line rides over all weaker sites. Consequently, in the stick-slide mode the height $H$ of the droplet is monotonous in time, whereas for the stick-jump mode it decreases during the pinning phase, but goes up at every jump.}

\begin{table}[ht]
\caption{Average initial contact angles $\theta_{00}$, contact angles just before ($\theta^*$), and after ($\theta_0$) the first observed jump, and the corresponding change in contact angle $\delta\theta=\theta_0-\theta^*$. The excess free energy $\delta\widetilde{G}$ per unit length of the contact line at the first jump. \textcolor{black}{Reported values represent averages of 6 and 2 measurements on hydrophobized silicon and glass substrates, respectively, which possess different roughnesses. The indicated deviations represent the observed variation of each value in different experiments}. As a comparison, the contact angle $\theta_{water-air}$ of a sessile water droplet ($L\approx1$mm) in air on both substrates is given.}
  \begin{tabular}{ |c | c | c | c | c | c | c |c |}
\hline
Surface 	&  $\theta_{00}$ 			&  $\theta^*$ 			& $\theta_0$			 & $\delta\theta$			 & rms roughness 	& $\theta_{water-air}$ 	& $\delta\widetilde{G}$\\
\hline
    Silicon 	& $48^{\circ} \pm4^{\circ}$	&  $41^{\circ}\pm1^{\circ}$ 	& $43^{\circ} \pm 2^{\circ}$	 & $2^{\circ}\pm1^{\circ}$ 	& 1.3 nm 		& $109^{\circ}$		& $7.9\times10^{-9}$ J/m\\
    Glass 	& $55^{\circ}\pm5^{\circ}$ 	&  $15^{\circ} \pm 5^{\circ}$     & $18^{\circ} \pm1^{\circ}$  	 & $3^{\circ}\pm2^{\circ}$ 	& 3.9 nm		& $107^{\circ}$ 		& $2.5\times10^{-7}$ J/m\\
    \hline
  \end{tabular}
\label{table1}
\end{table}

\section{Theory}

\subsection{Droplet dissolution}
The dissolution or evaporation of a droplet is driven by a negative concentration gradient of the droplet's constituent(s) from the droplet interface towards the surrounding medium. The case of a free spherical \textcolor{black}{air bubble in water} was solved by Epstein and Plesset \cite{epstein1950}. Their result has to be modified for sessile droplets, to reflect the modified geometry and the boundary conditions (i.e., no flux through the substrate). This was done \textcolor{black}{for evaporating droplets in air} by Popov \cite{Popov2005} in 2005:
\begin{equation}
\frac{dV}{dt} =  \frac{-\pi L D(c_s-c_{\infty})}{2\rho} f(\theta)
\label{dVdtF}
\end{equation}
with
\begin{equation}
f(\theta) = \frac{\sin(\theta)}{1+\cos(\theta)}+4\int_0^\infty \frac{1+\cosh(2\theta\epsilon)}{\sinh(2\pi\epsilon)}\tanh[(\pi-\theta)\epsilon]\mathrm{d}\epsilon
\end{equation}
and $\rho$ the density of the droplet material, $D$ the diffusion constant, and $L$ the \textcolor{black}{base} diameter of the droplet. $c_s$ is the concentration at the droplet-bulk interface, which equals the maximum solubility. Together with the concentration $c_{\infty}$ far away from the droplet, the solubility determines the undersaturation of the system,
\begin{equation}
\zeta = 1-\frac{c_{\infty}}{c_s}
\end{equation}
In the current system, $c_{\infty}=0$ and thus $\zeta=1$, which means that we assume the water far away from the droplet to be pure. This assumption is justified since after complete dissolution of our \textcolor{black}{alcohol} droplet, the concentration is only $\approx 10$ ppm of the maximum concentration.
Equation (\ref{dVdtF}) demonstrates that two droplets of equal volume can have distinct life times, depending on their geometry. The droplet life time is defined as the time from the start of the experiment, when the droplet has initial volume $V_{0}$, till complete dissolution. In the CA mode, where $\dot{\theta}=0$, equation (\ref{dVdtF}) can be rewritten to find $\dot{L}$. Integrating $\dot{L}$ from $L_0$ till $0$ then gives the life time. In the CR mode, where $\dot{L}=0$, the life time can be obtained by rewriting equation (\ref{dVdtF}) to find $\dot{\theta}$ and integrating from $\theta_0$ till $0$.  Stauber {\it{et al.}}\cite{Stauber2014} followed this route \textcolor{black}{for evaporating droplets in air }to obtain expressions for life times in the CA and CR modes, and combined the two for the stick-slide mode. They obtained the life time in the stick-slide mode as
\begin{equation}
\widetilde{\tau}_{SS} = \left(\frac{2\left(1+\cos\theta_0\right)^2}{\sin\theta_0 \left(2+\cos\theta_0\right)}\right)^{\frac{2}{3}}\left[\int_{\theta^*}^{\theta_0} \frac{2\mathrm{d}\theta}{f(\theta)\left(1+\cos\theta\right)^2} +\frac{\sin\theta^*\left(2+\cos\theta^*\right)}{f(\theta^*)\left(1+\cos\theta^*\right)^2} \right]
\label{tss}
\end{equation}
where the time is non-dimensionalized by
\begin{equation}
T=\left(\frac{3V_0}{2\pi}\right)^{\frac{2}{3}} \frac{\rho}{2D(c_{s}-c_{\infty})} 
\label{T}
\end{equation}
i.e., $\widetilde{\tau}=\tau/T$. The first term inside the square brackets in equation (\ref{tss}) is an integration over $\theta$ and reflects the \textcolor{black}{evaporation} in the CR mode. This mode is maintained until the contact angle has decreased to $\theta^*$, after which \textcolor{black}{evaporation} continues with a fixed contact angle $\theta^*$, reflected in the second term inside the brackets. By the choice of the timescale $T$, dimensionless life times $\widetilde{\tau}_{ss}$ between $0$ and $1$ are obtained, regardless the initial droplet size or material properties. \textcolor{black}{By using this scaling and the values for 3-heptanol in water, the above can be used to describe the present case.}

In the stick-jump mode, the slide phase is replaced by a quick jump. As indicated in the introduction, the duration of the jump is very short. We can therefore neglect the mass loss during the jump and the dissolution solely takes place in the CR mode. Referring to  Figure \ref{fig:sketch}, the total life time will thus be $\tau_{life}=\tau_1+\tau_3+\tau_5+$... To find all individual contributions, we follow the same approach as in equation (\ref{tss}) and integrate $\dot{\theta}$ between $\theta_0$ and $\theta^*$. In the stick-jump mode the \textcolor{black}{base} diameter changes during each jump, changing the dissolution rate. Therefore, $\tau_1> \tau_2>\tau_3$, and an additional term is required,
\begin{multline}
\widetilde{\tau}_{life} = \left(\frac{2\left(1+\cos\theta_0\right)^2}{\sin\theta_0 \left(2+\cos\theta_0\right)}\right)^{\frac{2}{3}}
\Bigg[\left(\int_{\theta^*}^{\theta_0} \frac{2\mathrm{d}\theta}{f(\theta)\left(1+\cos\theta\right)^2} \right) \\
\left(1-\left(\frac{\sin \theta^* \left(2+\cos \theta^*\right)}{\left(1+\cos \theta^*\right)^2}  \frac{\left(1+\cos\theta_0\right)^2}{\sin\theta_0\left(2+\cos\theta_0\right)}\right)^{\frac{2}{3}}\right)^{-1}\Bigg]
\label{tlife}
\end{multline}
The first term inside the square brackets is again the integration over the contact angle. The second term is the summation over all intervals. \textcolor{black}{A derivation of equation (\ref{tlife}) is given in appendix A.} 

In Figure \ref{fig:Lifetimes}, $\widetilde{\tau}_{life}$ is plotted as a function of $\theta_0$ for several values of $\theta^*$. For comparison, the life times in the CR, the CA and the stick-slide mode are also shown. Figure \ref{fig:Lifetimes} nicely illustrates that the life time in the stick-jump mode can exceed that in the stick-slide mode with as much as $10\%$. Only for $\theta_0>90^{\circ}$, the life time in the stick-slide mode can exceed that in the stick-jump mode, as shown in the inset in Figure \ref{fig:Lifetimes}.

\begin{figure}
\begin{center}
\includegraphics[angle=0,width=8cm]{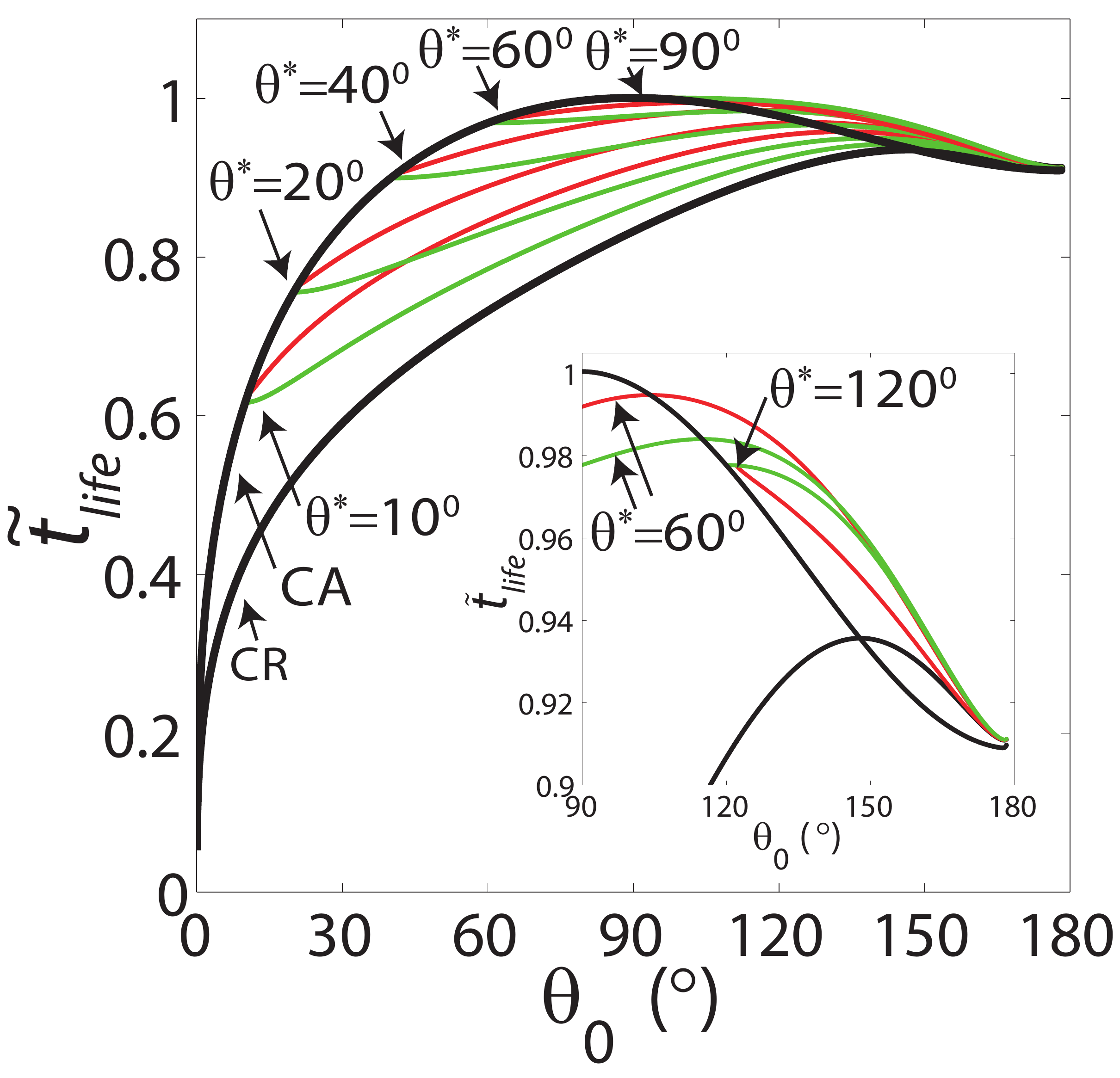}
\end{center}
\caption{\label{fig:Lifetimes} (color online) Scaled life times for droplets dissolving in the stick-slide (green curves), stick-jump (red curves) and the CA and CR modes (solid black curves), as a function of the initial contact angle. The inset shows that for $\theta_0>90^{\circ}$, the stick-slide and stick-jump mode life times are not bound by the two principal modes. For some combinations for $\theta_0$ and $\theta^*$, the life time of the stick-slide mode exceeds that of the stick-jump mode.}
\end{figure}

\subsection{Origin of the 'stick'}
The CR, the stick-slide, and the stick-jump modes require the contact line to be stationary during (parts of) the dissolution process. Experimentally we observe that directly after formation of the droplet, it has a certain contact angle that we refer to as $\theta _{00}$. When $\theta_{00}$ is larger than the receding contact angle $\theta^*$, the contact line does not move. Dissolution in the CR mode decreases the contact angle and when the contact angle reaches $\theta^*$, the contact line will start to move and the first jump is observed. Contact angle hysteresis can be caused by heterogeneities at the interface \cite{Gao2006} which can be of chemical nature, where 'patches' or islands with different hydrophobicity lead to local pinning of the contact line. Geometrical features can have a similar effect, for example when small pits, scratches, or bumps are present. 
Heterogeneities can originate from the droplet as well: The presence of micro- or nanoparticles inside the droplet will strongly alter the dissolution \textcolor{black}{or evaporation} dynamics. Convection inside the droplet will transport and deposit the particles near the contact line, giving rise to the well known coffee stain effect \cite{Marin2011}. The deposited particles then act as pinning sites and many researchers have shown that addition of nanoparticles can affect the stick-slide or stick-jump process \cite{Orejon2011, Maheshwari2008, Deegan2000}.
If the contact line is pinned during dissolution, the contact angle will increasingly deviate from the equilibrium value, resulting in an excess free energy. Shanahan \cite{Shanahan1995} presented a theory in which he models the pinning of the contact line as an energy barrier $U$. In the ideal case where $\theta_0$ is the equilibrium contact angle, and the droplet jumps between $\theta_0$ and $\theta^*$, the maximum excess free energy $\delta\widetilde{G}$ was derived by Shanahan as
\begin{equation}
\delta\widetilde{G} =  \frac{\gamma L \left(\delta\theta\right)^2}{4\left(2+\cos\theta_{0}\right)},
\label{dG}
\end{equation}
with $\gamma$ the interfacial tension of the droplet-bulk interface and $\delta\theta=\theta_0-\theta^*$ (see Figure \ref{fig:sketch}). The contact line stays pinned until $\delta\widetilde{G}$ exceeds the pinning energy barrier. The droplet then jumps back to $\theta_{0}$, with an associated change $\delta L$ in contact diameter. Equation (\ref{dG}) can be modified to express $\delta\widetilde{G}$ in terms of $\delta L$ instead of $\delta\theta$. By doing so, and assuming the potential barrier $U$ to be fixed during dissolution \textcolor{black}{or evaporation of the droplet,} Shananhan showed that $\delta L\propto\sqrt{L}$.  The feature of the small droplets as compared to larger ones becomes clear here: since $\delta L\propto\sqrt{L}$, we have $\delta L /  L\propto1/ \sqrt{L}$ and thus the relative jumps in contact line diameter are  larger when the droplet is small, making them easier to observe.

The work of Shanahan was recently commented on by Oksuz and Erbil \cite{Oksuz2014}. They explained how to calculate $\delta\widetilde{G}$ in experiments with an initial contact angle larger than the contact angle after the jump, e.g., when $\theta_{00}\neq\theta_0$, like in the present case. They argued that for the first observed jump, $\theta_{00}$ should be choosen as equilibrium angle, and $\theta_{00}-\theta^*$ should be taken as the deviation from equilibrium instead of $\delta\theta$. By following this and by using the value of $L$ just before the jump, as Oksuz and Erbil did, we can calculate the excess free energy of the contact line $\delta\widetilde{G}$ associated with the first jump for droplets on silicon and glass substrates. These values are listed in Table \ref{table1}. The value of $\delta\widetilde{G}=2.5\times10^{-7}$ J/m on glass is comparable to the values found by others \textcolor{black}{for evaporating droplets in air} \cite{Orejon2011}. The weaker pinning on silicon ($\delta\widetilde{G}\approx8\times10^{-9}$ J/m) corresponds to small deviations from the equilibrium contact angle, and tiny jumps.

\section{Comparison between experiment and theory}

\begin{figure}
\begin{center}
\includegraphics[angle=0,width=8cm]{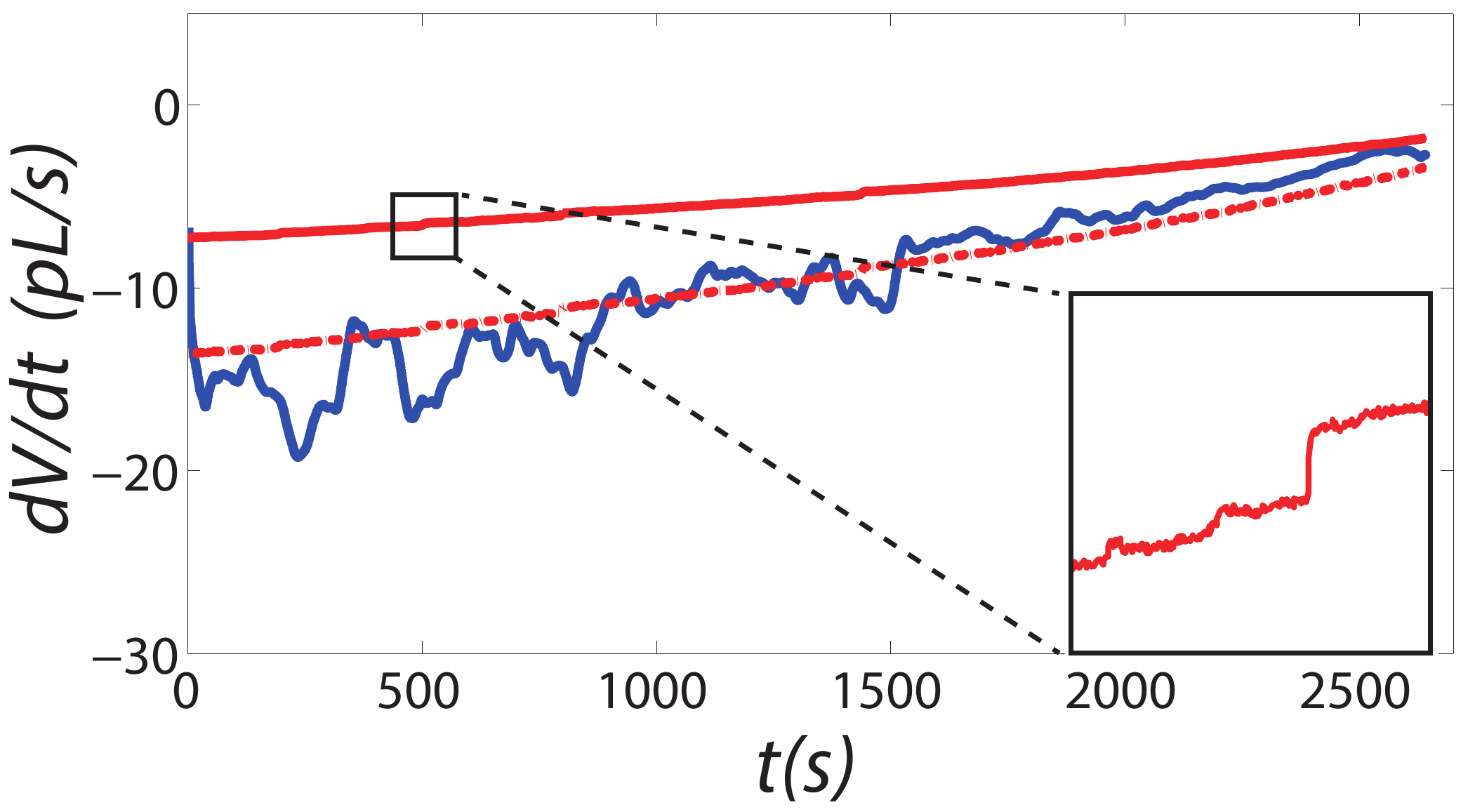}
\end{center}
\caption{\label{fig:dMdt} (color online) Actual volume loss as calculated from the frame-to-frame change in volume (blue line) and the expected dissolution rate (red). D$=0.8\times10^{-9}$ m$^2$s$^{-1}$ is used for the solid red line, the dashed line uses D$=1.6\times10^{-9}$ m$^2$s$^{-1}$, which is estimated from our measurements. The experimentally measured \textcolor{black}{base} diameter and contact angle are used as input variables in equation (\ref{dVdtF}) to calculate the red curve.}
\end{figure}

From equation (\ref{dVdtF}) we learn that the volume loss rate, and thus the life time, depends on the \textcolor{black}{base} diameter and contact angle. These values change during the experiment and from jump to jump. We therefore compare the expected volume loss per time unit, as calculated by equation (\ref{dVdtF}), to the actually measured volume change. This is done in Figure \ref{fig:dMdt}. The values for $L$ and $\theta$ at each instance in time are not fixed, nor predicted, but taken from the actual measurement. The calculated value underestimates the actually measured volume loss by a factor of $\approx 2$. Keeping in mind that we used an estimated value for the diffusion coefficient, we use Figure \ref{fig:dMdt} to estimate the actual diffusion constant in this system to be $D=1.6\times10^{-9}$ m$^2$s$^{-1}$. The calculation using this value is shown as the dotted red line in the same figure. The inset in Figure \ref{fig:dMdt} shows a zoom at the moment of a jump, which illustrates that the calculated dissolution rate before the jump is higher than after the jump.
There are multiple explanations for the difference in the theoretical and observed rate of dissolution. Firstly, the theoretical diffusion constant of $D \approx 0.8\times10^{-9}$m$^2$s$^{-1}$ is the value for 1-heptanol. The 3-heptanol molecule has a slightly different size, and its polar OH group is located at a different position, which might affect the diffusion constant. The exact influence is far from trivial, however, probably the difference will not be more than a few percent \cite{Hao1996}. Furthermore, equation (\ref{dVdtF}) neglects convection, which in case of temperature differences or air flow in the laboratory may be an issue in evaporation experiments. The present work has been conducted in water which provides better thermal stability and dampens convection because of the higher viscosity. Still, the dissolving droplet itself could induce natural convection in the water, or water evaporation from the surface of the tank could induce a flow inside the bulk. Convection will bring clean water in to the vicinity of the droplet, increasing the dissolution rate.\textcolor{black}{\cite{Poesio2009} Evidence for the existence of such a convection field, induced by the dissolving droplet itself, has recently been observed in preliminary experiments using particle image velocimetry.}

In Table \ref{table2}, the experimentally observed life times are compared to the theoretical values, as calculated using equation (\ref{tlife}). \textcolor{black}{Using $D=1.6\times10^{-9}$ m$^2$s$^{-1}$ and $\theta_0$ and $\theta^*$ as observed in each individual measurement, all data points are close to, or within $20\%$ of the expected value.} Perfect agreement is not obtained, which is not surprising since our model assumes perfect pinning during the 'stick' phase, a contact angle that alternates between two fixed values, and in addition a round contact area, which is not exactly the case, due to the inhomogeneous pinning sites. 
Better agreement is likely to be obtained in systems in which the pinning is perfect and facilitates well defined contact angles, like for example the concentric circles used by Debuisson and coworkers \cite{Debuisson2011a,Debuisson2011b}.

\begin{table}[ht]
\caption{Non-dimensionalized experimentally observed ($\widetilde{t}_{exp}$) and calculated ($\widetilde{t}_{life}$) life times for droplets on hydrophobized silicon and glass substrates. We used $D = 1.6\times10^{-9}$ m$^2$s$^{-1}$, as found earlier.}
  \begin{tabular}{|c|c|c|c|c|}
\hline
Experiment & Surface & \, $\widetilde{\tau}_{exp}$ \,& \, \,$\widetilde{\tau}_{life}$\,\, &${\widetilde{\tau}_{exp}}/\widetilde{\tau}_{life}$ \\
\hline
1 & Silicon &  $0.91\pm0.03$  &  $0.91\pm0.01$  & $1$  	 \\
2 & Silicon &  $0.80\pm0.03$  &  $0.91\pm0.01$  & $0.88$  	\\
3 & Silicon &  $0.90\pm0.03$  &  $0.92\pm0.01$  & $0.98$ 	 \\
4 & Silicon &  $0.84\pm0.03$  &  $0.92\pm0.01$  & $0.91$ 	\\
5 & Silicon &  $0.90\pm0.03$  &  $0.90\pm0.01$  & $1$	 \\
6 & Silicon &  $0.80\pm0.03$  &  $0.90\pm0.01$  & $0.89$ 	\\
7 & Glass  &  $0.85\pm0.03$  &  $0.70\pm0.01$  & $1.21$	\\
8 & Glass  &  $0.77\pm0.03$  &  $0.74\pm0.01$  & $1.04$	\\    
\hline
\end{tabular}
\label{table2}
\end{table}

\section{Conclusion}
In addition to the three known modi that have been described earlier, we have shown that a fourth mode, called the 'stick-jump' mode, can occur. The stick-jump mode could be studied in detail by using small dissolving droplets of 3-heptanol in water. We have shown that the dissolution of these droplets can be described by the same equations that apply to evaporating droplets. 

Characteristic in the stick-jump mode is intermittent contact line pinning, which we related to surface roughness. Geometric heterogeneities provide anchoring points for the contact line, and by increasing the roughness of the substrate, pinning was enhanced drastically, resulting in more pronounced stick-jump behaviour. We also explained why the stick-jump behavior is more easily observed for the small droplets used in this study, as compared to large ones.  

Experimentally observed life times of the droplets were in good agreement with the presented theoretical model. For initial droplet contact angles $\theta_0<90^{\circ}$, the life time in the stick-jump mode exceeds those in the constant radius mode or the stick-slide mode for initial contact angles. Even better agreement between measured and calculated life times might be obtained in future work, in which patterned substrates provide well-defined pinning sites. 

%\begin{suppinfo}
%Firstly, we give a more elaborate derivation of equation (\ref{tlife}). Secondly, detailed information on the surfaces is given. AFM images of the substrates before %and after coating are shown, and a description of the coating process is given.
%\end{suppinfo}

%%%%%%%%

\appendix
\section{Appendix A: Life time in stick-jump mode}

The rate of volume loss of a sessile droplet is given by equation (\ref{dVdtF}). The  geometric relation 
\begin{equation}
V = \frac{\pi L^3}{24}\frac{\sin\theta\left(2+\cos\theta\right)}{\left(1+\cos\theta\right)^2}
\label{eq2}
\end{equation} 
can be used together with equation (\ref{dVdtF}) to find an expression for $\frac{dL}{dt}$ to describe a droplet dissolving in the constant angle mode. In the constant radius mode, a similar approach can be used to find the change of $\theta$ in time:
\begin{equation}
\frac{d\theta}{dt} =  \frac{-4D(c_s -c_{\infty})}{\rho L^2}f(\theta)\left(\cos\theta+1\right)^2
\label{eq3}
\end{equation}
Equation (\ref{eq3}) can be integrated over $\theta$ from $\theta_0$ till $0$ to find the life time in the constant radius mode.

In the stick-jump mode, the contact line is pinned until the contact angle reaches the value $\theta^*$, which is when the line depins and jumps. The time required for a droplet with \textcolor{black}{base} diameter $L_0$ to go from $\theta_0$ to $\theta^*$ can be found by rewriting and integrating equation (\ref{eq3}):
\begin{equation}
\tau =  \frac{\rho L_{0}^2}{4D(c_s -c_{\infty})} \int_{\theta^*}^{\theta_0} \frac{d\theta}{ f(\theta)\left(\cos\theta+1\right)^2}
\label{eq4}
\end{equation}
After this time the droplet jumps to $\theta=\theta_0$, a new \textcolor{black}{base} diameter $L_1$, and the process starts over again. Assuming that the droplet always jumps from $\theta^*$ to $\theta_0$, the total life time of the droplet will consist of the combination of all intervals between the jumps:
\begin{multline}
\tau_{life} =  \frac{\rho}{4D(c_s -c_{\infty})} \biggl(L_0^2    \int_{\theta^*}^{\theta_0} \frac{d\theta}{ f(\theta)\left(\cos\theta+1\right)^2}+\\
L_1^2  \int_{\theta^*}^{\theta_0} \frac{d\theta}{ f(\theta)\left(\cos\theta+1\right)^2} +L_2^2  \int_{\theta^*}^{\theta_0} \frac{d\theta}{ f(\theta)\left(\cos\theta+1\right)^2}+..... \biggr)
\label{eq5}
\end{multline}
with $L_0$ the initial \textcolor{black}{base} diameter, $L_1$ the diameter after the first jump, $L_2$ after the second jump, etc. 

The initial volume $V_0$ relates to the initial \textcolor{black}{base} diameter $L_0$ and the initial contact angle $\theta_0$ via: 
\begin{equation}
V_0 = \frac{\pi L_0^3}{24}\frac{\sin\theta_0\left(2+\cos\theta_0\right)}{\left(1+\cos\theta_0\right)^2}
\label{eq6}
\end{equation}
Just before the first jump, when $\theta=\theta^*$ but still $L=L_0$, this volume has decreased to $V_{0,end}$:
\begin{equation}
V_{0,end} = \frac{\pi L_0^3}{24}\frac{\sin\theta^*\left(2+\cos\theta^*\right)}{\left(1+\cos\theta^*\right)^2}
\label{eq7}
\end{equation}
Since the volume is assumed to be conserved during the jump, $V_{0,end} = V_1$, which is equivalent to:%%%
%%%%
\begin{equation}
V_1 = \frac{\pi L_1^3}{24}\frac{\sin\theta_0\left(2+\cos\theta_0\right)}{\left(1+\cos\theta_0\right)^2} = \frac{\pi L_0^3}{24}\frac{\sin\theta^*\left(2+\cos\theta^*\right)}{\left(1+\cos\theta^*\right)^2}
\label{eq8}
\end{equation}
which can be re-arranged to find
\begin{equation}
L_1^3= L_0^3\frac{\sin\theta^*\left(2+\cos\theta^*\right)}{\left(1+\cos\theta^*\right)^2}\frac{\left(1+\cos\theta_0\right)^2} {\sin\theta_0\left(2+\cos\theta_0\right)}
\label{eq8a}
\end{equation}
The same can be done around the second jump, when $V_{1,end}=V_2$:
\begin{equation}
V_{1,end} = \frac{\pi L_1^3}{24}\frac{\sin\theta^*\left(2+\cos\theta^*\right)}{\left(1+\cos\theta^*\right)^2} = \frac{\pi L_2^3}{24}\frac{\sin\theta_0\left(2+\cos\theta_0\right)}{\left(1+\cos\theta_0\right)^2}
\label{eq8b}
\end{equation}
Substituting equation (\ref{eq8a}) in to (\ref{eq8b}) results in an expression for $L_2$ in terms of $L_0$, $\theta_0$ and $\theta^*$:
\begin{equation}
L_2^3= L_0^3\left(\frac{\sin\theta^*\left(2+\cos\theta^*\right)}{\left(1+\cos\theta^*\right)^2}\right)^2\left(\frac{\left(1+\cos\theta_0\right)^2} {\sin\theta_0\left(2+\cos\theta_0\right)}\right)^2
\label{eq8c}
\end{equation}
This procedure can be repeated to find a general expression for the \textcolor{black}{base} diameter of the droplet after the N-th jump:
\begin{equation}
L_N^3={L_0^3}\left(\frac{\sin\theta^*\left(2+\cos\theta^*\right)}{\left(1+\cos\theta^*\right)^2}\right)^N
\left(\frac{\left(1+\cos\theta_0\right)^2}{\sin\theta_0(2+\cos\theta_0)}\right)^N
\label{eq8d}
\end{equation}
Equations (\ref{eq4}) and (\ref{eq8d}) can be combined to obtain the life time of a droplet dissolving in the stick-jump mode:
\begin{equation}
\tau_{life} =  \frac{\rho L_0^2}{4D(c_s -c_{\infty})}\left[\int_{\theta^*}^{\theta_0} \frac{\mathrm{d}\theta}{f(\theta)\left(1+\cos\theta\right)^2} \right] \sum_{N=0}^{\infty} \left(\frac{\sin \theta^* \left(2+\cos \theta^*\right)}{\left(1+\cos \theta^*\right)^2}  \frac{\left(1+\cos\theta_0\right)^2}{\sin\theta_0\left(2+\cos\theta_0\right)}\right)^{\frac{2N}{3}}
\label{eq9}
\end{equation}
which can be simplified to 
\begin{equation}
\tau_{life} =  \frac{\rho L_0^2}{4D(c_s -c_{\infty})}
\left[\int_{\theta^*}^{\theta_0} \frac{\mathrm{d}\theta}{f(\theta)\left(1+\cos\theta\right)^2} \right] \\
\left(1-\left(\frac{\sin \theta^* \left(2+\cos \theta^*\right)}{\left(1+\cos \theta^*\right)^2}  \frac{\left(1+\cos\theta_0\right)^2}{\sin\theta_0\left(2+\cos\theta_0\right)}\right)^{\frac{2}{3}}\right)^{-1}
\label{eq9a}
\end{equation}

As derived by Picknett and Bexon \cite{Picknett1977} and Stauber {\it{et al.}} \cite{Stauber2014}, the maximum life time that a droplet can achieve is in the constant angle mode with $\theta=90^{\circ}$. Stauber {\it{et al.}} non-dimensionalized the life time $\tau_{life}$ in such a way that the scaled life time ($\widetilde{\tau}_{life}$) in this situation equals 1 \cite{Stauber2014}, i.e., with $T$ as in equation (\ref{T}).
 Scaling in this way allows for easy comparison between the different modes. Adopting the same non-dimensionalization to our expression results in:
\begin{multline}
\widetilde{\tau}_{life} = \left(\frac{2\left(1+\cos\theta_0\right)^2}{\sin\theta_0 \left(2+\cos\theta_0\right)}\right)^{\frac{2}{3}}
\left[\int_{\theta^*}^{\theta_0} \frac{2\mathrm{d}\theta}{f(\theta)\left(1+\cos\theta\right)^2} \right] \\
\left(1-\left(\frac{\sin \theta^* \left(2+\cos \theta^*\right)}{\left(1+\cos \theta^*\right)^2}  \frac{\left(1+\cos\theta_0\right)^2}{\sin\theta_0\left(2+\cos\theta_0\right)}\right)^{\frac{2}{3}}\right)^{-1}
\label{eq11}
\end{multline}

\section{Appendix B: Substrate roughness}

\begin{figure}
\begin{center}
\includegraphics[angle=0,width=8cm]{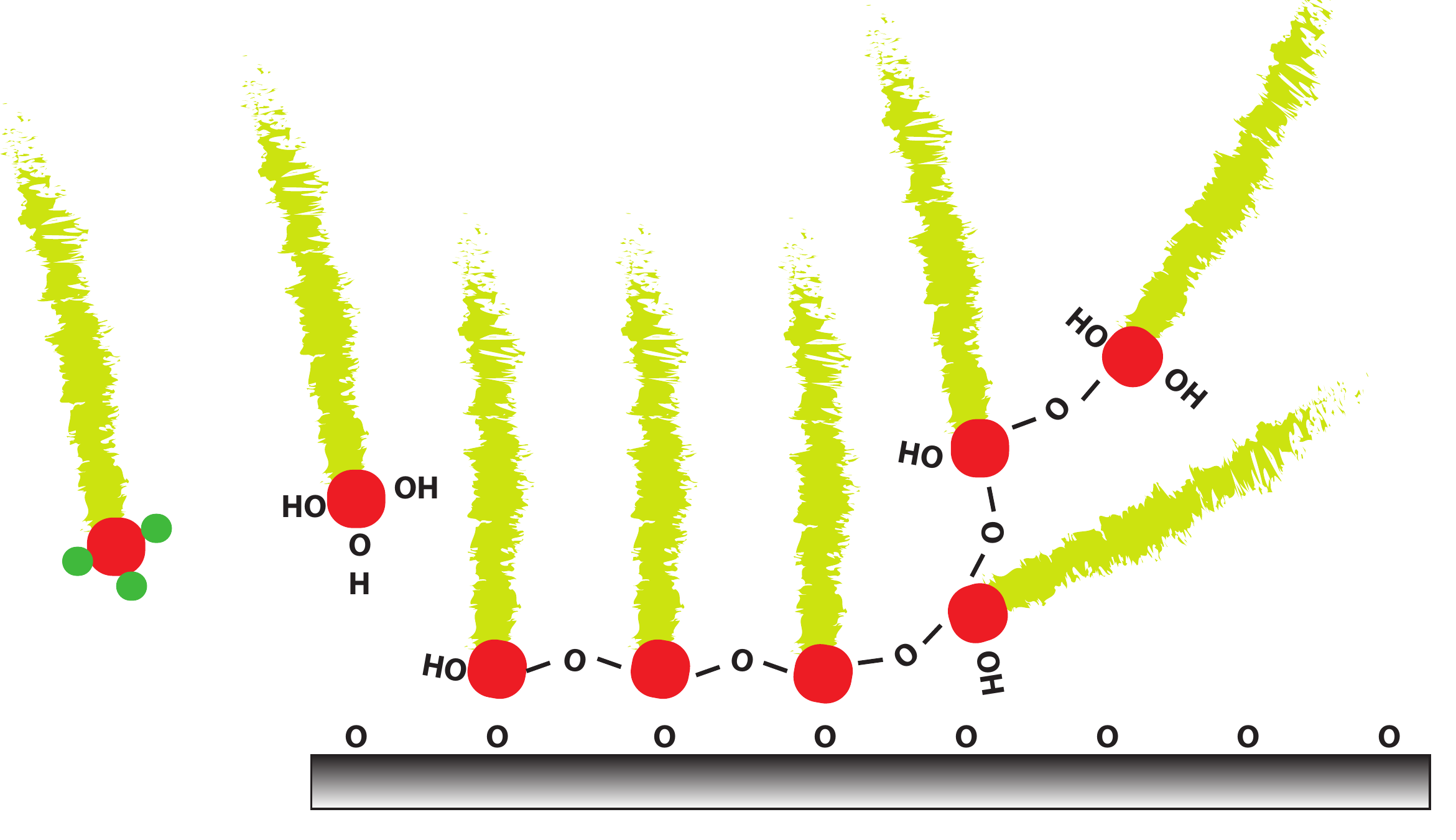}
\end{center}
\caption{\label{PFDTS} (color online)PFDTS molecules consist of a hydrophobic tail (yellow), a silicon head (red) with three chlorine atoms. The chlorine then reacts with water to form HCl, and the silicon head is hydroxilated. Middle three molecules illustrate the idealy desired situation, where the headgroups are attached to the interface and their neighbors. However, the molecules can also polymerize and form superstructures , as indicated by the rightmost molecules. }
\end{figure}

Two substrates were used in this study, namely glass microscopy slides and silicon wafers. The silicon wafers are crystalline and polished, providing an almost atomically smooth substrate. Glass microscopy slides are made from soda-lime glass, which is an amorphous material, and is known to posses a roughness of a few nanometers \cite{North2010}. Both substrates have a surface that consists of SiO$_2$ groups, which can be coated by chloro-silanes. A self assembled monolayer of such silanes will result in a hydrophobic, chemically homogeneous surface \cite{Schwartz2001}.
The chloro-silane used in this work (1H,1H,2H,2H-perfluorodecyltrichlorosilane, or PFDTS) consists of a fluorinated carbon tail and a silicon headgroup with three chlorine atoms. During vapor deposition, which is illustrated in Figure \ref{PFDTS}, the head group reacts with some residual water that is present in the reactor, replacing the chlorine with OH groups. These OH groups initially bind to the OH groups at the substrate surface via hydrogen bonds. Further polymerization produces water, and results in very strong covalent Si-O-Si bonds. Ideally, one of the three OH groups of the silane head bonds to the substrate, while the others polymerize with neighboring molecules, forming a dense and strong monolayer. In practice, it is observed that these molecules can polymerize into cross-linked superstructures \cite{Wang2005}, illustrated in the right part of Figure \ref{PFDTS}. 

\begin{figure}
\begin{center}
\includegraphics[angle=0,width=8cm]{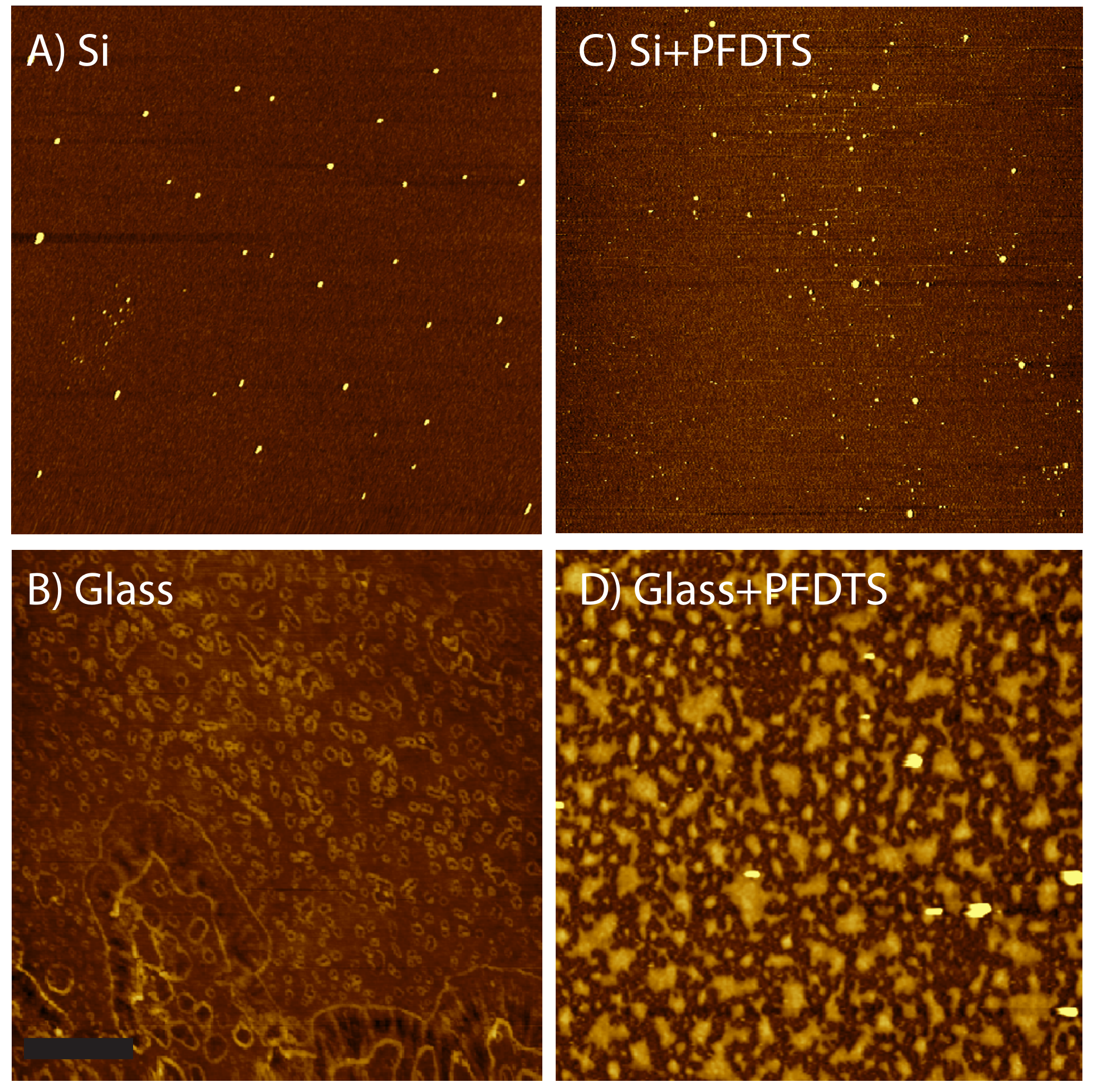}
\end{center}
\caption{\label{Substrates} (color online) AFM images of uncoated silicon (A) and glass substrates (B). Coating these surfaces with PFDTS slightly increases the roughness of the silicon substrate (C), whereas polymerization of the molecules around the crevices in glass leads to the formation of geometrical heterogeneities. Height scales are $5$ nm (A+C), $10$ nm (B) and $30$ nm (D), respectively. Scan sizes are $5\times5$ $\mu$m$^2$, and the scalebar represents $1$ $\mu$m. Panels A-C and B-D do not show the same surface area.  }
\end{figure}

The growth of silane superstructures  affects the substrate appearance. Figure \ref{Substrates} shows AFM images of the untreated silicon (A), and glass (B) surfaces. Apart from some small particles, uncoated silicon has a smooth surface. The glass substrate has an intricate pattern of ridges and pits, which are most probably formed during the production. 
After coating, some small particles are visible on the the silicon substrate in Figure \ref{Substrates}C, however, the roughness of the substrate has not been changed significantly by the coating. This is contrasted by the glass substrate in Figure \ref{Substrates}D which features many large protrusions. We hypothesize that these protrusions are aggregates of polymerized PFDTS. Earlier work has shown that increased substrate roughness can cause PFDTS molecules to polymerize and form superstructures on its own \cite{AkramRaza2010}. This undesired polymerization is enhanced when crevices are present in the substrate. Water can remain inside these hydrophilic crevices during vapor deposition, and will accelerate the polymerization of PFDTS molecules upon contact. Our hypothesis that the observed features are polymerized PFDTS is supported by the fact that the size and coverage of the protrusions are comparable to those of the crevices in the uncoated substrate. Cleaning the coated substrate in toluene or chloroform, which are solvents for unbound PFDTS, does not alter the surface appearance, indicating that the molecules inside the aggregates are chemically bonded to each other and to the surface.

%%%FIGURE%%%%%%%%%%%%%%%%%%%%%%%%%%%%%%%%%%%%%%%%%%
%%%FIGURE%%%%%%%%%%%%%%%%%%%%%%%%%%%%%%%%%%%%%%%%%%
%%%FIGURE%%%%%%%%%%%%%%%%%%%%%%%%%%%%%%%%%%%%%%%%%%
\newpage

\bibliography{JumpBib}
%%%%%%%%
\end{document}